\documentstyle[12pt]{article}
\input{psfig}
\def\be{\begin{equation}}
\def\ee{\end{equation}}
\def\bea{\begin{eqnarray}}
\def\eea{\end{eqnarray}}
\newcommand{\gsim}{\lower.7ex\hbox{$\;\stackrel{\textstyle>}{\sim}\;$}}
\newcommand{\lsim}{\lower.7ex\hbox{$\;\stackrel{\textstyle<}{\sim}\;$}}
\def\tw{\theta_{\!\rm w\,}}
\def\sintw{\sin\tw}
\def\costw{\cos\tw}
\def\bm#1{{\mbox{\boldmath $#1$}}}
\newcommand{\phat}[1]{\hat{\phi}^{#1}}

\newcommand{\eps}{\epsilon}
\newcommand{\Fem}{{\cal F}^{\rm em}}

\newcommand{\Aem}{A^{\rm em}}

\begin{document}
\footnotesep=14pt
\begin{flushright}
\baselineskip=14pt
{\normalsize DAMTP-1999-25}\\
{\normalsize DAMTP-1999-26}\\
{\normalsize {hep-ph/9902432}}
\end{flushright}

\vspace*{.5cm}
\renewcommand{\thefootnote}{\fnsymbol{footnote}}
\addtocounter{footnote}{1}
\begin{center}
{\Large\bf
Magnetic Fields from Bubble Collisions ---}\\*[1ex]
{\Large\bf A Progress Report\footnote{\noindent Summary of
presentations given at the International Conference
on Strong and Electroweak Matter (SEWM-98),
Copenhagen, Denmark, 2-5 Dec 1998, and at the
19th Texas Symposium on Relativistic Astrophysics, Paris,
14-18 Dec 1998, to appear in both proceedings.}}
\end{center}
\begin{center}
\baselineskip=16pt
{\bf Ola T\"{o}rnkvist}\footnote{E-mail:
{\tt o.tornkvist@damtp.cam.ac.uk}}\\
\vspace{0.4cm}
{\em Department of Applied Mathematics and Theoretical Physics,}\\
{\em University of Cambridge,}\\
{\em Silver Street, Cambridge CB3 9EW,
United Kingdom}\\
\vspace*{0.25cm}{15 February 1999}
\end{center}
\baselineskip=20pt
\vspace*{.5cm}
\begin{quote}
\begin{center}
{\bf\large Abstract}
\end{center}
\vspace{0.2cm}
{\baselineskip=10pt Primordial magnetic fields may be created
in collisions of expanding bubbles in a first-order cosmological
phase transition and later serve as seed fields for galactic
magnetic fields. Here I present results from
numerical and analytical studies of U(1) bubble collisions
done in collaboration with Ed Copeland and Paul Saffin.
I also reveal preliminary results from analytical studies of
SU(2)$\times$U(1) bubble collisions, which provide first evidence that
magnetic fields are created in
electroweak two-bubble \mbox{collisions}.}
\end{quote}
\renewcommand{\thefootnote}{\arabic{footnote}}
\setcounter{footnote}{0}
\newpage
\baselineskip=16pt
\section{Introduction}
Observations show that magnetic field strengths of the order
of $10^{-6}$ Gauss are typical
in spiral galaxies. Because the fields appear aligned with the rotational
plane of each galaxy, a plausible explanation is that they were created
by a dynamo mechanism, in which a weak seed field was
exponentially amplified by the turbulent motion of ionised gas
in conjunction with
the differential rotation of the galaxy. There have been many proposals
for seed fields \cite{lackofspace};  here
I shall concentrate on the generation of magnetic fields in bubble
collisions in the electroweak phase transition (EWPT).
Although there is no
consensus that such seed fields would be sufficiently strong and
correlated \cite{Son}, recent results indicate that the present theoretical
bounds on seed fields are too stringent \cite{LTD}.

Because there is no EWPT
in the Standard Model (SM) \cite{Kajantie}, we need to consider
a two-Higgs model such as
the Minimal Supersymmetric Standard Model (MSSM), which
permits a strong first-order transition
for Higgs-boson mass $M_h\lsim 105-116$
GeV \cite{LaineCline}. For the purpose of bubble
nucleation a more weakly first-order transition will
suffice,\footnote{The
bounds on $M_h$ quoted in Ref.~\cite{LaineCline} were obtained by requiring
that $\langle\phi(T_{\rm c})\rangle/T_{\rm c}>1$, which corresponds to
a first-order phase transition strong enough also to
prevent wash-out of the baryon asymmetry produced in electroweak
baryogenesis.}
and
consequently higher values of $M_h$ are allowed. A general
(non-supersymmetric) two-Higgs model can easily accommodate a
first-order transition within its extensive parameter space.

Because two-Higgs models
and the SM have the same Higgs
vacuum manifold $S^3$, bubble collisions in these models
are expected to be very similar, and we can
focus our attention on the more tractable SM problem. In these
investigations
we ignore plasma friction and the dissipation due to Ohmic currents.

\section{Bubble Collisions in a U(1) Model}
Let us first consider a U(1) model and
the collision of two bubbles with Higgs fields
$\phi_1=\rho_1(x) e^{i\theta_1}$ and $\phi_2=\rho_2(x) e^{i\theta_2}$,
respectively, and constant phases
$\theta_1\neq\theta_2$.
In order to model the generation of magnetic fields, we
set the initial U(1) field strength to zero. One may then choose
a gauge in which the vector potential $V_\mu$ is initially zero.
Because of the phase gradient, a gauge-invariant current $j_k =
iq[\phi^{\dagger}D_k\phi - (D_k \phi)^{\dagger}\phi]$ develops across
the surface of intersection of the two bubbles, where $D_k=\partial_k
-iqV_k$. The current gives rise to a ring-like flux of the field
strength
$F_{ij}=\partial_i V_j - \partial_j V_i$ which takes the shape of a girdle
encircling the bubble intersection region. The time evolution of
the azimuthal field strength, as obtained in recent numerical
simulations \cite{us}, is shown in Fig.~1b.
An
approximate
time-dependent solution was first obtained by Kibble and
Vilenkin \cite{KibVil}, who
set
the modulus of the Higgs field
equal to
its vacuum expectation value. Because they used a step
function as initial condition for the phase of the Higgs field, their
solution suffers from discontinuities on the future null surface of
bubble intersections.

We have obtained a different analytical solution
by taking into account the
profile of the Higgs modulus for each bubble and deriving smooth
initial conditions from a superposition ansatz \cite{us}. These
analytical
results can be seen in Fig.~1c and may be compared with the exact
evolution in Fig.~1b.
The magnetic field is
concentrated in a narrow flux tube along the circle of most recent
intersection of the two bubbles. From the analytical solutions we
find that the tube's radial width is
$\Delta r\sim m^{-1} t_{\rm
coll}/t$ and the maximal field strength is $B_{\rm max} \sim
|\theta_2-\theta_1| m^2 r(t)/(2qt_{\rm coll})$,
where $t_{\rm coll}$
is the time of collision, $r(t)$ is the bubble radius, and $m$ is the
vector-boson mass. The total flux tends to a constant
$(\theta_2-\theta_1)/q$ at large times \cite{KibVil}.
The narrowing of the flux tube
and increase of field strength with time is a result of Lorentz
contraction as the bubble walls accelerate in the absence of friction.

\section{Magnetic Fields from SU(2)$\times$U(1) Bubble Collisions\/}

In electroweak bubble collisions the initial Higgs field in the
two bubbles is of the form
\be
\label{higgs}
\Phi_1 = 
\Big(\!\begin{array}{c} 0\\[-1mm]\rho_1(x)\end{array}
\!\Big)~,
\quad\quad
\Phi_2 = \exp(i\textstyle{\theta_0} \bm{n}\cdot\bm{\tau})
\Big(\!\begin{array}{c} 0\\[-1mm]\rho_2(x)\end{array}
\!\Big)~,
\ee
respectively. A constant SU(2) matrix
$\exp(i\textstyle{\theta_0} \bm{n}\cdot\bm{\tau})$ has here
replaced the phase
$e^{i\theta_0}$ of the U(1) model.
Saffin and Copeland have found that
the initial Higgs-field configuration
can be written globally as\,\cite{edpaul}
\be
\Phi(x)= \exp(i\textstyle{\theta(x)} \bm{n}\cdot\bm{\tau})
\Big(\!\begin{array}{c} 0\\[-1mm]\rho(x)\end{array}
\!\Big)~,
\label{globh}
\ee
where
$\bm{n}$
is the same constant unit vector as in Eq.~(\ref{higgs}).
As the bubbles collide,
non-Abelian currents $j^A_k=i [\Phi^{\dagger} T^A D_k\Phi -
(D_k\Phi)^{\dagger} T^A \Phi]$ develop
across the surface of intersection of the two bubbles, where
$T^A=(g'/2, g \tau^a/2)$, $D_k=\partial_k-iW_k^AT^A$ and
$W^A_k=(Y_k,W^a_k)$.
In analogy with the U(1) case, one obtains here a ringlike flux,
but this time of non-Abelian field strengths.
The important question is then how to project out the
electromagnetic
field in a gauge-invariant way. In recent work \cite{bdef}
I have shown
how to construct from first principles
an electromagnetic U(1) vector potential, unique up to a gradient,
whose curl {\em in any gauge of
SU(2)$\times$U(1)\/} gives
the electromagnetic (EM) field tensor.
In terms of the
3-component unit isovector $\hat{\phi}^a=
(\Phi^{\dagger}\tau^a\Phi)/(\Phi^{\dagger}\Phi)$ this vector
potential is given by\footnote{This expression
is actually regular as
$\bm{\hat{\phi}}\to \bm{s}$,
which can be seen
e.g.\ by
expressing $\bm{\hat{\phi}}$ in spherical coordinates \cite{bdef}.}
\be
\Aem_{\mu}=-\sintw\phat{a} W^a_\mu +\costw Y_\mu
-\frac{\sintw}{g}
\frac{\eps^{abc}s^a\phat{b}\partial_\mu\phat{c}}{1-s^d\phat{d}}~,
\label{adef}
\ee
where $\bm{s}=\{s^a\}$ is an arbitrary constant unit vector.
It is not hard to show that $\Aem_\mu$ transforms only by a pure gradient
under
general SU(2)$\times$U(1) gauge transformations
(or a change of
$\bm{s}$). The line integral of $\Aem_\mu$ along a closed curve
therefore gives the gauge-invariant EM flux. The
gauge-invariant EM field tensor $\Fem_{\mu\nu}\equiv\partial_\mu\Aem_\nu
-\partial_\nu\Aem_\mu$ becomes \cite{bdef}
\be
{\cal F}^{\rm em}_{\mu\nu} =
-\sintw \partial_{[\mu}(\phat{a} W^a_{\nu]}) + \costw F^Y_{\mu\nu}
+ \textstyle\frac{\sintw}{g} \epsilon^{abc} \phat{a}
(\partial_\mu\phat{})^b (\partial_\nu\phat{})^c~.
\label{Fem}
\ee
It is important to note that $\Fem_{\mu\nu}$ receives contributions from
Higgs-field gradients. The benefits
of definition (\ref{Fem}) as compared with
other proposed definitions have been discussed thoroughly in
Refs.~9 and 10.

We now return to the issue of bubble collisions.
As in the U(1) case one may assume that the initial configuration
has zero field strengths and zero vector potentials,
$F^a_{\mu\nu}=F^Y_{\mu\nu}=W_\mu^a=Y_\mu=0$.
The only possible contribution to the EM field tensor
then comes from the last term in Eq.~(\ref{Fem}). However,
for $\phat{a}$ corresponding to
the initial Higgs configuration
Eq.~(\ref{globh}),
this term is zero \cite{eworigin}.
The simple reason is that
 $\phat{a}$ depends on the
spacetime coordinates $x$ only through {\em one\/} scalar function
$\theta(x)$, and no antisymmetric tensor can be constructed
from its derivatives.
Similarly I have shown that,
as long as the field configuration maintains the simple form (\ref{globh}),
the electric current vanishes \cite{eworigin}.
{}From Maxwell's equations one
then finds that the time derivatives of both electric and magnetic
fields are zero.

It would then seem as though no magnetic fields are generated in
electroweak two-bubble collisions. This is indeed what I find
using the {\em linearised\/} equations to calculate
analytically the field evolution beyond initial times.
However,
there is a magnetic
field proportional to $n_3(1-n_3^{\,\,2})\sin^3\tw/g$
generated to first order in the nonlinearities by a current produced
by the initially excited W and Z fields. The current arises
because of the difference in the evolution of the W and Z mass eigenstates.
This first-order correction is furthermore
characterised by a rotational motion in the $(W^1,W^2)$ isoplane.
The resulting current generates
a magnetic field to second order in the nonlinearities.

By contrast, in a collision of three bubbles
a magnetic field is present already in the
initial field configuration. Labelling the bubbles by 0, 1, and 2, we
denote the Higgs field in each bubble by $\Phi_i=U_i (0,\rho_i(x))^{\top}$,
where $U_0=1$, $U_1=\exp(i\theta_1\bm{m}\cdot\bm{\tau})$, and
$U_2=\exp(i\theta_2\bm{n}\cdot\bm{\tau})$ are constant SU(2) matrices with
linearly independent unit vectors\pagebreak[4]
$\bm{m}$, $\bm{n}$. The initial
Higgs field can here be written globally as
$\Phi=\exp[i(f(x) m_a + g(x) n_a)\tau^a] (0,\rho(x))^{\top}$,
where
$f,g = \sin\theta_{1,2}\rho_{1,2}\Theta/(\rho\sin\Theta)$,
$\cos\Theta = \sum_i\rho_i\cos\theta_i/\rho$
and $\rho^2=\sum_i\rho_i^2 +
2\sum_{i<j}\rho_i\rho_j(
\cos\theta_i\cos\theta_j+\bm{m}\cdot\bm{n}\sin\theta_i\sin\theta_j)$
with $\theta_0=0$.
{}From Eq.~(\ref{Fem})
the initial electromagnetic field then becomes
\be
{\cal F}^{\rm em}_{\mu\nu} = \textstyle
-\frac{4\sintw}{g}  f_{[,\mu}g_{,\nu]}\,\,\hat{\bm{e}}_3\cdot
\!\left(\frac{\sin  2\Theta}{2\Theta} (\bm{m}\!\times\!\bm{n}) +
\frac{\sin^2\Theta}{\Theta^2} (f\bm{m}\!+\!g \bm{n})\!\times\!
(\bm{m}\!\times\!
\bm{n} )\right)~,
\label{threebub}
\ee
which comprises a nonzero magnetic field as long as
$\nabla f\times\nabla g\neq 0$. The three-bubble mechanism is,
however, suppressed,
as the third bubble must impinge before the SU(2)
phases of the first two have equilibrated, and such nearly simultaneous
collisions are rare.

If $\bm{m}$ and $\bm{n}$ are collinear, ${\cal F}^{\rm em}_{\mu\nu}$ vanishes.
More generally, one can show that the EM field and
the current are zero only
when the SU(2) phases of the Higgs field take
values along one single geodesic on the Higgs vacuum manifold,
such as is generated by the exponentiation of one Lie-algebra element
(cf.\ Eq.~(\ref{globh})).

\begin{figure}[p]
\psfig{figure=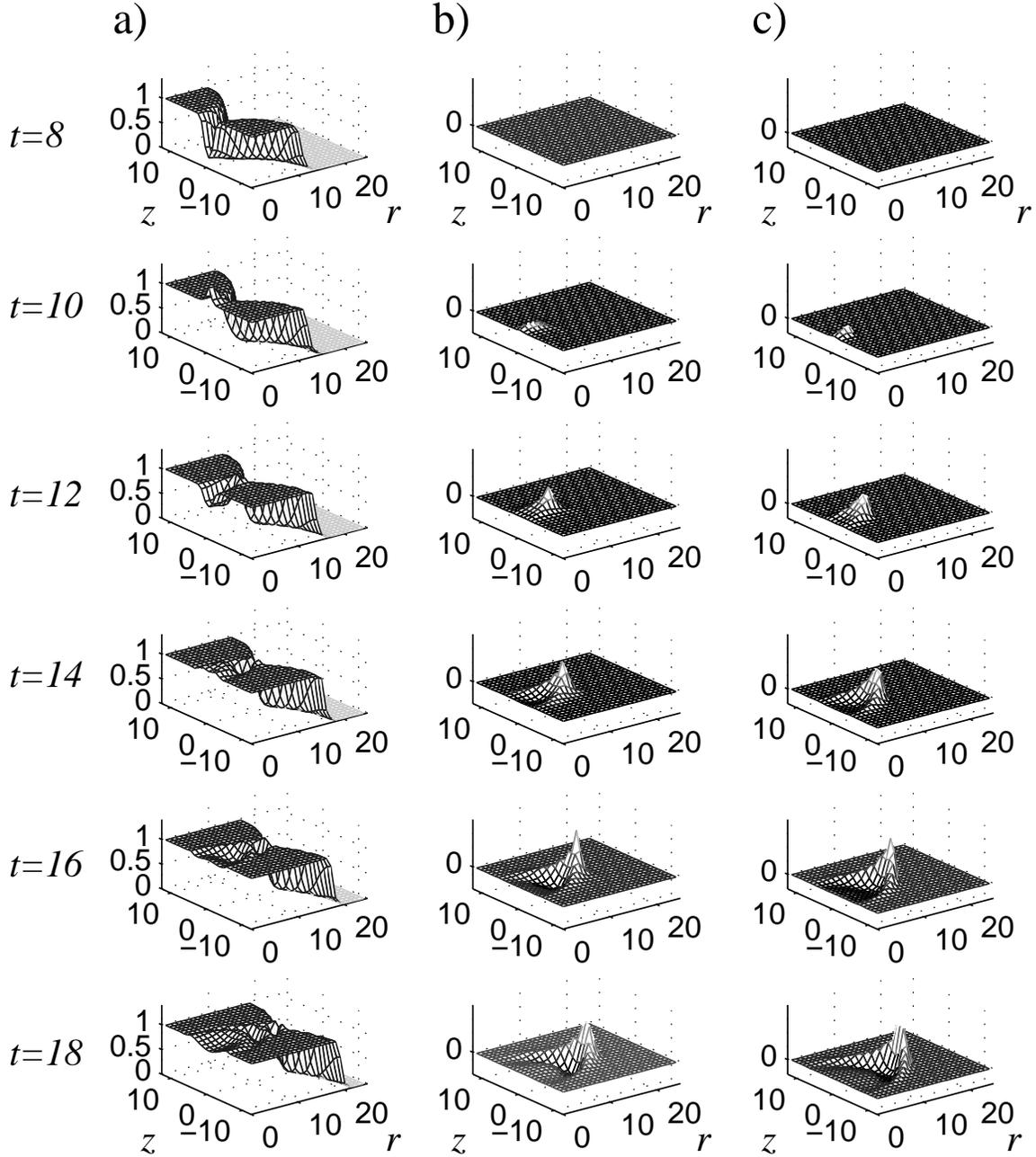,height=18.0cm}
\caption{Bubble collision in a U(1) model: Evolution of
a) modulus of the Higgs
field, b) the azimuthal field strength $F_{rz}$
in a numerical
simulation. c) Analytical solution.
Here $r=(x^2+y^2)^{1/2}$
is the distance from the axis of the collision,
and $r$, $z$, $t$ are measured in units of $2 M_h^{-1}$. The centres of
the two bubbles are at $r=0$, $z=\pm 12.5$. The collision occurs shortly
before $t=10$. A colour version of this figure can be viewed and
downloaded at {\tt http://www.damtp.cam.ac.uk/user/ot202/sewm98/}
  \label{fig:bubs}}
\end{figure}

\section*{Acknowledgments}
The author is supported
by the European Commission's TMR programme under Contract
No.~ERBFMBI-CT97-2697.


\begin{thebibliography}{99}
\bibitem{lackofspace}Due to lack of space here, please see references given
in [10] below.

\bibitem{Son} D.T.\ Son, hep-ph/9803412.

\bibitem{LTD} M.J.\ Lilley, O.\ T\"ornkvist, and A.-C.\ Davis,
in preparation.

\bibitem{Kajantie} K.\ Kajantie, 
M.\ Laine, K.\ Rummukainen and
M.\ Shaposhnikov,
{\em Phys.\ Rev.\ Lett.\ }{\bf 77}, 2887 (1996);
F.\ Csikor, Z.\ Fodor and J.\ Heitger,
{\em Phys.\ Rev.\ Lett.\ }{\bf 82},
21 (1999).

\bibitem{LaineCline}
M.\ Laine and K.\ Rummukainen, {\em Phys.\ Rev.\ Lett.\ }
{\bf 80} 5259 (1998); J.M.\ Cline and G.D.\ Moore,
{\em Phys.\ Rev.\ Lett.\ }{\bf 81}, 3315 (1998).

\bibitem{us} E.J.\ Copeland, P.M.\ Saffin, and O.\ T\"ornkvist,
in preparation.

\bibitem{KibVil} T.W.B.\ Kibble, A.\ Vilenkin, {\em Phys.\ Rev.\ }{\bf D52},
679 (1995).

\bibitem{edpaul} P.M.\ Saffin and E.J.\ Copeland,
{\em Phys.\,Rev.\,} {\bf D56}, {1215} (1997).

\bibitem{bdef} O.\ T\"ornkvist, hep-ph/9805255.

\bibitem{eworigin} O.\ T\"ornkvist, {\em Phys.\,Rev.\,}{\bf D58},
043501
(1998).

\end{thebibliography}
\end{document}